# What is fair to ask Society to fund ? *

Orfeu Bertolami

Instituto Superior Técnico
Departamento de Física
Av. Rovisco Pais, 1
1049-001 Lisboa
Portugal

E-mail: orfeu@cosmos.ist.utl.pt
(http://alfa.ist.utl.pt/~orfeu/homeorfeu.html)

## Abstract

Our world is strongly driven by technological developments that continuously create new markets and shape individual taste and choice. The scale of dissemination of the most recent technological breakthroughs and the capability of creating them is what divides wealthy societies from poor ones. The increasing sophistication of these developments make it harder and harder for developing countries to actively contribute to these breakthroughs as well as to acquire the right to benefit from them through the exchange of products obtained through low-tech agriculture, industry and services.

Thus, as new technological developments rely often on discoveries in fundamental science and technology that have their origins, in some instances, many decades ago, it is only through fostering the acquisition of the skills necessary for innovation in basic science and technology that economical and social progress can be insured on a long term basis. In this perspective, funding of research is an investment for the future and clearly, just buying technology is no short cut to create sustainable wealth.

Furthermore, Society should fund fundamental research for cultural and training reasons. Indeed, as a representative of a country or of a culture in a given historical period, activities assertive of cultural and national identities should be funded. These involve necessarily scientific developments too. Moreover, experience shows that the best training for innovation is acquired from individuals and institutions that have an active interest in basic science and on its application to technology.

Why should Society fund research that brings no visible technological benefit ?
There is no straightforward answer to this question, but past experience has taught us to expect the unexpected. We have learned that discoveries which changed our civilization were made while answering questions and issues that were meaningful only in the specfic context they were posed. Some well known examples come to mind:

1. Michael Faraday discovered the electrical engine in 1821 and subsequently the electromagnetic induction in 1831, while searching for what was then a purely theoretical conjecture, namely the connection between electricity and magnetism.

2. Telecommunications have developed from the discovery of the electromagnetic waves by Heinrich Hertz in 1885, however he actually aimed to confirm the mathematical theory of the electromagnetic fields proposed by James Clerk Maxwell in 1864.

3. The basic circuits of computers were invented in the 1930s by nuclear physicists eager to tackle specific problems involving nuclear particles.

4. Nuclear power was not the product of applied scientists working on new energy sources, but the result of fundamental discoveries about the nature of matter by the Curies, Ernest Rutherford, Enrico Fermi and others in early XX century.

5. Transistors were invented by Bill Shockley, Walter Brattain and John Bardeen in 1948 based on theoretical ideas of Shockley and Bardeen, experts in the quantum theory of solids.

6. The so-called "green revolution" has emerged from the continous evolution of ideas and concepts in the fundamental science of plant genetics.

7. Tim Berners-Lee created the World Wide Web in 1990 as a by-product of his research on how to connect pieces of information from different computers, a task required for the treatment of the massive amount of data in particle physics.

Many others could be mentioned. Nevertheless, researchers are under continuous pressure to justify their work on practical terms. This is not always possible and somehow not even desirable given that the benefits of research on fundamental science is, most often, noticeable much after scientific discoveries have taken place. A surprising example of forsightness is due to Faraday, who when questioned over the esotherical nature of his research and the use of electricity by William Gladstone, a leading figure of conservative politics in the second half of the XIX century in Britain, emphatically replied: "One day, Sir, you may tax it".  On the other side, it could be remembered that Rutherford has always remained sceptical about the possibility of ever exploiting nuclear energy.

It is appropriate to remind the pragmatism of J.J. Thomson, Nobel Prize of Physics in 1906 for the discovery of the electron, in what concerns the difference between research in applied science and in pure science:

"A research on the lines of applied science would doubtless have led to improvement and development of the older methods – the research in pure science has given us an entirely new and much more powerful method. In fact, research in applied science leads to reform, research in pure science leads to revolutions, and revolutions whether political or industrial, are exceedingly profitable things if you are on the winning side."

In favour of research on basic science, we would like to further argue that the increasing degree of sophistication of nowdays technological developments are such that they require a quite especialized working force and rather specific training. It is common knowledge that the best training for scientific and technological innovation is acquired in institutions and from individuals that have a good record in innovative thinking and active interest in science and technology. This is even more evident given the fact that many technological devices incorporate discoveries in technology and basic science that, in some instances, took place many decades ago. Therefore, buying technology is not a satisfactory answer for long term sustainable wealth. Moreover, since the Industrial Revolution technology has become increasingly valuable against products that are manufactured through low-tech agriculture, industry and services and clearly it is more and more difficult to achieve innovation without a serious educational and intellectual involvement. Therefore, it is only through active policies favouring acquisition of especialized knowledge over a long span of time that satisfactory levels of success can be achieved on economical terms. For these reasons it is no wonder that the recruitment of especialized working force is an active policy of the most developed countries, and most noticebly by the United States. These policies are not restricted to importing "brains", but are extended also to the use of especialized skills in multinational companies and via tele-working.

Thus, an inescapable implication of the above line of reasoning is that the support and funding of higher education institutions and laboratories with good record on innovation must be stable and independent of political changes. Moreover, it is clear that the success of these institutions in promoting scientific and technological progress requires that they are universally accessible on the basis of merit, free from market pressures and preferentially funded by public sources. Needless to say that these institutions must be protected from ideologies and policies according to which knowledge and education are goods like any other in the market. Education is a fundamental human right that must be protected from the evolution and rules of the market.

The third reason that makes us believe that research in fundamental science and in technological developments with no visible spin-offs should be funded is because they are also assertive of an important cultural tradition of our civilization.

For sure, the outcome of most cultural activities cannot be measured for their value in the market:

What is the market value of a symphony ? Or a poem ? Or a theorem in Algebraic Topology ? Or a model in Theoretical Cosmology ?

It is self evident that these activities will never be profitable and that they cannot be judged by their impact on the material life of the consumers. In spite of that, it is somewhat consensual that they should, when intrinsically original, be minimally supported as they reflect the evolution of national identity and culture. They represent the quintessential contribution of a nation and of an historical period to the human civilization. The continuation of this cultural tradition is beyond question regardless the economical and market pressures. It should be also pointed out that science has provided us with unique means for better judging our standing in the whole picture of the Cosmos and for more profundly undertanding the responsabilities we should have with respect to humankind, to our most immediate environment and to the planet.

Finally, we would like to reflect on the importance of international scientific and technological collaboration. The perpetuation throughout the XXI century of the existing divisions in the present world is growing factor of instability. For how long will it be acceptable to deny to most of humankind the benefits of the latest developments in medicine, agricultural science, telecommunications and so on ? For how long will it remain acceptable turning the scientific edge into militar might and technology that can be exported and sold as any merchandise ? The inexistence of consensual solutions and global political strategies to ease up the building tension leaves us with little to hope for on the short term. International scientific and technological collaboration being no panacea, is at least a way of facilitating the growth of deeper interconnections and cooperation. International scientific and technological collaboration makes concrete the exercise of sharing information, knowledge and experience. It makes possible that past mistakes are avoided and that development can be really thought on global terms. Science and culture will be, as they were in the past, instrumental in improving our world and drawing new horizons for humankind. It is through the universality of the values emerging from science that new international institutions will eventually arise to prevent and cure the evils we see in our world as well as to meet the collective challenges humankind will enevitably face in the future.

It should be reminded that global collaborations have already reached an impressive degree of entanglement in some areas of research, and they exist in various degrees in virtually all areas of knowledge. Some of them are remarkable and most conspicuous. For example, in particle physics, the Counseil European pour la Research Nucleaire (CERN) is an umbrela of scientific activities in which hundreds of laboratories and universities around the world participate; in termonuclear fusion, a large fraction of the international scientific community (Canada, E.U., China, Japan, Russia, South Korea and U.S.A.) aims to construct the ITER, the International Thermonuclear Reactor, in the hope to achieve the control of thermonuclear fusion and to unleach an abundant and clean new source of energy. There are also various collaborations in astronomy such as the European Southern Observatory (ESO) and the Hubble Space Telescope Institute. In space science the European Space Agency (ESA) integrates the efforts of 15 european countries and the International Space Station represents a particularly important step in an area that was till the recent past very much associated to the Cold War.  In molecular biology, the European Molecular Biology Laboratory (EMBL), has more than just an european dimension and so does the Human Genome project. There are many areas where international collaborations have a signficant expression. Extending further these networks is, in our opinion, of capital importance for improving the  relationships between scientists and peoples of the world.  Continuously guaranteeing that funds are suitably directed to these collaborations is a major responsability of the international scientific community. Ensuring that developing countries get the most out of these collaborations should be regarded as a moral imperative for scientists of developing countries.

Parede, 20 October 2004.